# Visualization of Local Conductance in MoS$_2$/WSe$_2$ Heterostructure Transistors

Di Wu[1†], Wei Li[2†], Amritesh Rai[2], Xiaoyu Wu[1], Hema C. P. Movva[2], Maruthi N. Yogeesh[2], Zhaodong Chu[1], Sanjay K. Banerjee[2], Deji Akinwande[2], Keji Lai[1*]

[1] Department of Physics, The University of Texas at Austin, Austin, TX 78712, United States

[2] Microelectronics Research Center, Department of Electrical and Computer Engineering, The University of Texas at Austin, Austin, TX 78758, United States

[†] DW and WL contributed equally to this work.

* E-mail: kejilai@physics.utexas.edu

# Abstract

The vertical stacking of van der Waals (vdW) materials introduces a new degree of freedom to the research of two-dimensional (2D) systems. The interlayer coupling strongly influences the band structure of the heterostructures, resulting in novel properties that can be utilized for electronic and optoelectronic applications. Based on microwave microscopy studies, we report quantitative electrical imaging on gated molybdenum disulfide ($MoS_2$)/tungsten diselenide ($WSe_2$) heterostructure devices, which exhibit an intriguing antiambipolar effect in the transfer characteristics. Interestingly, in the region with significant source-drain current, electrons in the n-type $MoS_2$ and holes in the p-type $WSe_2$ segments are nearly balanced, whereas the heterostructure area is depleted of mobile charges. The spatial evolution of local conductance can be ascribed to the lateral band bending and formation of depletion regions along the line of $MoS_2$–heterostructure–$WSe_2$. Our work vividly demonstrates the microscopic origin of novel transport behaviors, which is important for the vibrant field of vdW heterojunction research.

Van der Waals (vdW) heterostructures (HSs) with atomically sharp interfaces and controllable layer components have been intensively investigated in the past few years[1-5]. The interlayer vdW coupling, although much weaker than the intralayer covalent bonding, is sufficient to modify the band structures of the assembled materials. For instance, when two disparate semiconducting transition-metal dichalcogenide (TMDC) layers are stacked together, the energy band offset[6] may strongly influence the carrier dynamics[7-16] and transport characteristics[17-27]. Due to the atomically thin nature of few-layer TMDCs, the depletion region in the vertical direction across the heterojunction, which is ubiquitously found in heterojunctions of three-dimensional (3D) semiconductors, does not exist in such 2D systems[19]. Taking advantage of these new features, researchers have demonstrated novel electronic and optoelectronic devices such as heterojunction photodiodes[17-20], resonant/Esaki tunneling diodes[21-23], multivalued logic inverters[24], electron-hole multiplication photocells[27], floating-gate memories[28], and excitonic transistors[29]. It is anticipated that the research in vdW heterostructures will continue to thrive in the foreseeable future.

Among the various TMDC heterostructures, the combination of $MoS_2$ and $WSe_2$ has received a lot of attention[9,10,13,15,17-19,21-26,28-30]. Due to the presence of S vacancies and strong metal Fermi-level pinning near its conduction band edge, $MoS_2$ usually exhibits n-type characteristics[31-33]. $WSe_2$, while being ambipolar, is mostly used as a p-type semiconductor because of its high hole mobility[34,35]. The subsequent formation of p-n junctions and the type–II band alignment with large band offsets[6] have led to interesting phenomena such as band-to-band tunneling[22,24,26], negative differential resistance[22,24,30], and ultra-long valley lifetime[15], among others. On the other hand, transport and most optical measurements are inherently macroscopic in nature and the sample response is averaged over large areas. In this work, we report the electrical mapping of a gated $MoS_2$/$WSe_2$ heterostructure transistor by microwave impedance microscopy (MIM)[36]. Transport across the device during the electrostatic gating shows a clear antiambipolar behavior[19,22-24], which was also observed in other heterostructure systems[37-39]. Interestingly, while the local conductance of individual $MoS_2$ and $WSe_2$ layers exhibits the usual n-type and p-type behavior, respectively, the heterostructure area is fully depleted of free carriers when an appreciable source-drain current can flow through the device. We show that such a counterintuitive spatial evolution of electrical conductance can be described by the band bending model and the presence of lateral depletion regions. Our work elucidates the underlying microscopic properties of vdW heterojunction devices, which is highly desirable for the continuous advance of 2D heterostructure research.

The TMDC HSs in this work were prepared by stacking exfoliated few-layer $MoS_2$ and $WSe_2$ flakes onto $SiO_2$ (300 nm)/Si substrates using a hot pick-up technique with polypropylene carbonate (PPC)-coated polydimethylsiloxane (PDMS) stamps[40,41]. After the assembly, the samples were annealed in vacuum at 300 °C to improve the interlayer coupling. Details of the HS preparation steps are found in the Methods section. Metal electrodes on the TMDC flakes, 20 nm Pd/ 30 nm Au on $WSe_2$ and 5 nm Ti/ 45 nm Au on $MoS_2$, were patterned by standard electron-beam lithography, electron-beam evaporation and lift-off techniques (see Methods section). Quasi-Ohmic contacts were formed at the metal-TMDC interface at room temperature, as confirmed by the linear current-voltage (I-V) characteristics on individual materials (Supporting Information S1). For the experiment presented here, we do not observe any dependence on the stacking order between $MoS_2$ and $WSe_2$.

Fig. 1a shows the optical image of a typical $MoS_2/WSe_2$ transistor device with metal contacts. As seen from the atomic-force microscopy image and the line profile in Fig. 1b, both the $MoS_2$ and $WSe_2$ flakes are around 2 nm in thickness, corresponding to 3 monolayers. The samples are sufficiently thin such that the depletion region in the vertical direction is negligible[24]. We therefore do not expect qualitative difference between our devices and thinner TMDC samples. The positions of the characteristic Raman peaks on individual flakes as well as the HS overlap region (Fig. 1c) are consistent with that reported in the literature[26,42,43]. The transfer and output characteristics of the HS device are plotted in Fig. 1d and 1e, respectively. As the back-gate voltage $V_{BG}$ increases from a very negative value of −20 V, the source-drain current ($I_{DS}$), starting from a negligible value, rises at $V_{BG} = -15$ V, peaks at around $V_{BG} = -10$ V, and drops back to the off state beyond $V_{BG} = 0$ V. In this work, such an unusual antiambipolar behavior is seen in all $MoS_2/WSe_2$ devices (Supporting Information S2). The same effect was reported in the literature[19,22-24] and has been attributed to the lateral tunneling of carriers across the HS transistor. However, the underlying spatial distribution of electrical conductance in such HS transistors has not been studied in previous reports. Finally, upon further increase of $V_{BG}$, $I_{DS}$ rises again since the current can now flow through the n-type $MoS_2$ and the n-branch of the ambipolar $WSe_2$ [23,24].

The gate-dependent local conductance of the $MoS_2/WSe_2$ transistor was studied by a microwave impedance microscope (MIM)[36] based on a tuning fork (TF) atomic-force microscope (AFM)[44,45]. As illustrated in Fig. 2a, an electrochemically etched tungsten tip with a diameter around 100 nm, which sets the spatial resolution of the technique, is attached to a quartz TF for topographic feedback. The MIM signal, modulated by the TF resonant frequency (~ 32 kHz), is demodulated by a lock-in amplifier to form the AC_MIM image. Details of the TF-based MIM and the driving-amplitude-modulation (DAM) feedback mode are found in Ref. 45. The sample thickness measured by the TF-AFM is the same as that determined by conventional contact-mode AFM (Supporting Information S3). During the experiment, the source and drain electrodes are grounded and the DC offset of the tip is set to be zero through a bias-tee such that the tip does not behave as a top gate on the sample.

In Fig. 2b, the $I_{DS}$-$V_{BG}$ curve around the antiambipolar peak at a small $V_{DS}$ of 0.3 V is replotted in the linear scale and divided into three distinct transport regimes (regions I, II and III corresponding to different $V_{BG}$ ranges). For simplicity, we only present selected AC_MIM-Im (proportional to the imaginary component of the tip-sample admittance) data, which are monotonic as a function of the local conductance (see analysis below) in Fig. 2c. The complete set of MIM images are included in Supporting Information S4. On individual TMDC flakes, the MIM signals gradually increase (decrease) with increasing $V_{BG}$ on the $MoS_2$ ($WSe_2$) part of the device. In contrast, the MIM signals on the HS overlap area display a non-monotonic gate dependence, which can be divided into three regions in accordance with the transfer curve of Fig. 2b. Being conductive at $V_{BG} = -20$ V, the HS gradually turns resistive with increasing $V_{BG}$ within Region I, starting from the $MoS_2$-HS boundary and moving inward. In the antiambipolar Region II ($-15$ V $< V_{BG} < -2.5$ V), the MIM signals on the HS area are essentially the same as that on the substrate, indicative of an insulating behavior. For further increase of $V_{BG}$ in Region III, conductivity reappears inside the HS, again starting from the $MoS_2$ side and moving toward the interior. The non-uniform MIM signals along the left edge of the HS area are presumably due to local defects and variation in the interlayer coupling. At a very positive $V_{BG} = 20$ V, the HS area is uniformly conductive except for a dark line on the right side, possibly due to the folding of flakes. Note that the MIM signal here is slightly lower than that on the highly conductive $MoS_2$. The same evolution of AC_MIM data are also observed in other $MoS_2$/$WSe_2$ devices in this study (Supporting Information S5).

For a quantitative understanding of the MIM data, we have performed finite-element analysis (FEA) of the tip-sample interaction[46] using commercial software COMSOL 4.4. Details of the geometric and electrical parameters for the simulation are shown in Supporting Information

S6. For the TF-based MIM, the tip-sample admittance is first simulated as a function of the tip height. As the tip taps on the sample surface, the admittance oscillates at the same TF resonant frequency. The first harmonic signal obtained by Fourier transform of the time-domain simulation curve then corresponds to the MIM data acquired by the lock-in amplifier. Fig. 3a shows the simulated AC_MIM signal with respect to the 2D sheet conductance $\sigma_{sh}$, which is indeed a monotonically increasing function in the relevant regime of $10^{-8}$ S/m $< \sigma_{sh} < 10^{-5}$ S/m. A comparison between the FEA result and the measured data in Fig. 3b allows us to quantitatively extract the average sheet conductance, as plotted in Fig. 3c. We emphasize that the gate dependence of local conductance in this HS device is nontrivial. In particular, the HS area is highly insulating in Region II where the trans-conductance is significant. The resistance between the source and drain electrodes is, therefore, not the summation of the resistances of the three segments, as is expected in a simple series circuit.

The counterintuitive local sheet conductance imaged by the MIM can be understood by the lateral band alignment across the device. With strong interlayer coupling, the overlap area of the HS device can be viewed as a new TMDC material, whose energy bands are hybridized from $MoS_2$ and $WSe_2$. While band bending in the vertical direction is absent in few-layer TMDC materials, the drift and diffusion of free carriers will still lead to band bending in the horizontal direction and the formation of depletion regions. The lateral band alignment across the three segments at zero source-drain bias is qualitatively illustrated in Fig. 4a. In Region I with highly negative $V_{BG}$, the $MoS_2$, HS, $WSe_2$ areas are depleted, weakly p-type, and strongly p-type, respectively. As $V_{BG}$ increases and the Fermi level $E_F$ rises, depletion region inside the HS first appears near the $MoS_2$ side and gradually moves towards the interior of the HS. Note that the lateral width of this insulating section is small at the left and right corners of the HS area, where the two individual TMDC layers are in close proximity. As the device is forward biased ($V_{DS} > 0$ V) and the potential barrier in the HS lowered, the recombination current will flow mainly through these corners rather than having to traverse the center of the HS. The resultant output characteristics are similar to that of the traditional p-n junction diode (see Fig. 1e), except that a significant forward current only occurs when both the p-$WSe_2$ and n-$MoS_2$ regions are sufficiently conductive around $V_{BG} = -12$ V. As $V_{BG}$ further increases, $MoS_2$ becomes heavily n-doped and $WSe_2$ is depleted. Carriers can no longer travel across the device and the transport current drops again. For even higher $V_{BG}$, it is expected that WSe$_2$ also becomes n-doped and $I_{DS}$ increases again. The fact that the WSe$_2$ region remains insulating at $V_{BG}$ = 20 V may be due to certain surface electrochemistry, which can be mitigated by a capping layer in future work. The evolution of local conductance in the middle of the HS area is vividly seen in the line profiles plotted in Fig. 4b. The gradual change of MIM signals along the line of MoS$_2$-HS-WSe$_2$ indicates that the band bending occurs in a length scale of 0.5 ~ 1 μm. Such information is not accessible through transport studies and can only be obtained by the electrical mapping in our MIM measurements.

In summary, we demonstrate the quantitative electrical imaging of few-layer MoS$_2$/WSe$_2$ heterostructure transistors by microwave impedance microscopy. The local sheet conductance on the n-type MoS$_2$ and p-type WSe$_2$ sections of the device changes monotonically as a function of the back-gate voltage. The heterostructure area, on the other hand, is depleted of free carriers in the middle of the antiambipolar region where the transport current can flow through the device. The gate dependence of local conductance evolution can be satisfactorily explained by the lateral band alignment model. Since nanoscale conductivity is an essential property for electronic applications, our spatial mapping of the local conductance is of crucial importance for the rapid progress of vdW heterostructure research.

## Methods

**Van der Waals heterostructure preparation:** Few-layer MoS$_2$ and WSe$_2$ flakes were first mechanically exfoliated onto SiO$_2$ (300 nm)/Si substrates using Scotch tape and Gel-Pak film, and then annealed at 300 °C in vacuum (base pressure ~ 10$^{-6}$ Torr) for 6 h to remove tape residues. The flake thicknesses were determined using atomic-force microscopy (AFM) measurements and were found to be ~ 2 nm for both MoS$_2$ and WSe$_2$. The heterostructures were assembled by stacking one MoS$_2$ flake onto another WSe$_2$ flake using a hot dry-transfer pick-up technique with polypropylene carbonate (PPC)-coated polydimethylsiloxane (PDMS) stamps. A flake of MoS$_2$ was first picked up by the PDMS/PPC stamp at 40 °C and then stacked onto another WSe$_2$ flake on the target SiO$_2$ (300 nm)/Si substrate at 110 °C. The high temperature enabled delamination and release of the PPC film onto the target substrate, and was critical to reduce the trapped moisture and gases in the heterostructure interface region. After the PPC was released, the target substrate was soaked in chloroform, acetone and isopropanol consecutively to remove the PPC residue. The heterostructure samples were then annealed in vacuum at 300 °C to improve the interface quality in order to enhance the interlayer coupling.

**Device fabrication and transport measurement:** Metal contacts were deposited via electron-beam evaporation (base pressure ~ 5 x $10^{-6}$ Torr) after the electrode regions were defined by conventional electron-beam lithography utilizing poly(methyl methacrylate) (PMMA) as the resist. Different metals were chosen to form low resistance quasi-Ohmic contacts with different TMDC flakes at room temperature: 20 nm Pd/30 nm Au was deposited as contacts for $WSe_2$ and 5 nm Ti/ 45 nm Au was deposited as contacts for $MoS_2$. After the contact lift-off in acetone, the surface of the heterostructure flakes was further cleaned by contact-mode AFM scanning before the MIM and transport measurements. The transport measurements were carried out in the dark using Keithley 4200 and 4201 Semiconductor Characterization Systems (SCS) under ambient conditions.

**MIM measurements:** A tuning-fork-based MIM was employed to map the local conductance. In particular, an electrochemically etched tungsten tip was attached to a quartz tuning fork (resonant frequency ~ 32 kHz). A Zurich HF2L1 lock-in amplifier was used to control the tuning-fork tip to vibrate at its resonant frequency in the driving amplitude modulation (DAM) mode[45]. The topography feedback system was provided by a commercial AFM system (Park XE-70). The AC_MIM signals were demodulated by a SR830 lock-in amplifier and then acquired by the Park AFM system. During the measurements, source and drain electrodes were grounded, and the back-gate voltage was applied using a Keithley 2400 Source Measure Unit (SMU) to modulate the carrier density.

**Finite-element analysis (FEA):** Finite-element analysis by COMSOL 4.4 was carried out to quantify the sheet conductance of TMDC flakes. Since the lateral dimensions of the flakes are much larger than the MIM tip diameter, the 2D axisymmetric model can be used here. Details of the geometric and electrical parameters are found in Supporting Information S4. We then follow the standard procedure in Ref. 45 to convert the demodulated tip-sample admittance to the AC_MIM output based on the calibration of our electronics. Note that the AC_MIM signal is plotted as a function of the 2D sheet conductance $\sigma_{sh} = \sigma_{3D} \cdot t$, where $\sigma_{3D}$ is the 3D conductivity and *t* is the flake thickness. Since *t* is much smaller than the tip diameter, the MIM response of $MoS_2$, HS, $WSe_2$ segments follows the same simulation curve shown in Fig. 3a.

## Acknowledgments

The MIM experiment was supported by the U.S. Department of Energy (DOE), Office of Science, Basic Energy Sciences, under the Awards No. DE-SC0010308 and DE-SC0019025. The sample fabrication was supported by the Welch Foundation Grant No. F-1814. D.A. acknowledges the Presidential Early Career Award for Scientists and Engineers (PECASE). This work was partly done at the Microelectronics Research Center (MRC) of the Texas Nanofabrication Facility, a member of the National Nanotechnology Coordinated Infrastructure (NNCI) supported by the National Science Foundation (NSF) grant ECCS-1542159 as well as the Multidisciplinary University Research Initiative (MURI) grant W911NF-17-1-0312. A.R. and S.K.B. would also like to acknowledge support from the NASCENT Engineering Research Center (ERC) funded by NSF grant EEC-1160494.

## Supporting Information

Additional details on transport and microwave imaging results of other heterostructure devices, as well as finite-element analysis. This material is available free of charge via the Internet at http://pubs.acs.org.

## Author Contributions

K.L. conceived the project. D.W. and W.L. fabricated the devices and performed the transport measurements. A.R., H.C.P.M. and M.N.Y. assisted with the heterostructure preparation. D.W. conducted the MIM experiments, performed data analysis and drafted the manuscript with K.L. All authors have contributed to the manuscript preparation and have given approval to the final version of the manuscript.

## Conflict of Interest

The authors declare no competing financial interest.

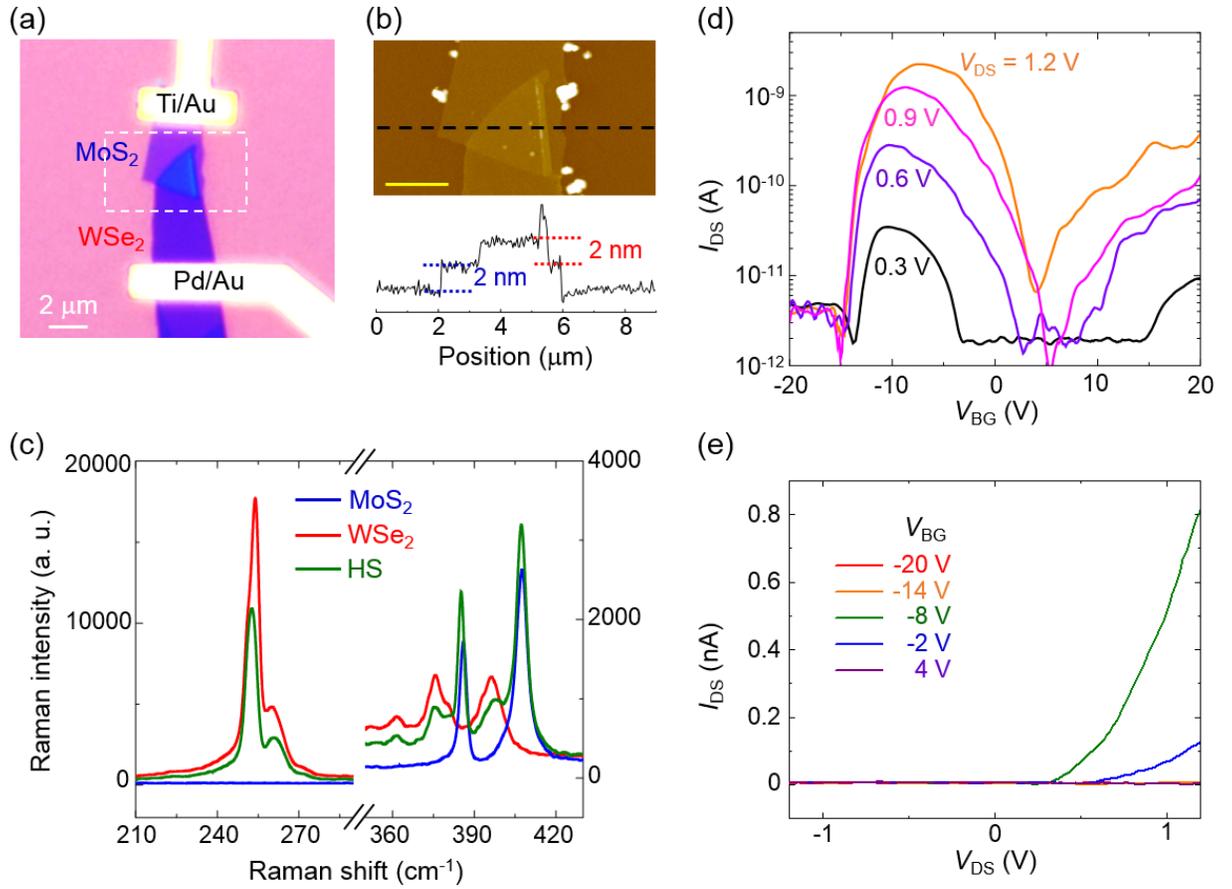

Figure 1. (a) Optical image of the $MoS_2$/$WSe_2$ heterostructure device with source (Ti/Au) and drain (Pd/Au) electrodes. (b) Top: AFM image inside the white dashed box in (a). The bright particles are tape residues. The scale bar is 2 μm. Bottom: line profile across the black dashed line in the AFM image, showing a thickness of 2 nm for both flakes. (c) Raman spectra acquired on three different locations of the sample with 532 nm excitation line. The heterostructure (HS) region displays characteristic Raman peaks of both $MoS_2$ and $WSe_2$. (d) Transfer characteristics of the HS transistor device under several source-drain biases. (e) Output characteristics of the device in (d) under several back-gate voltages.

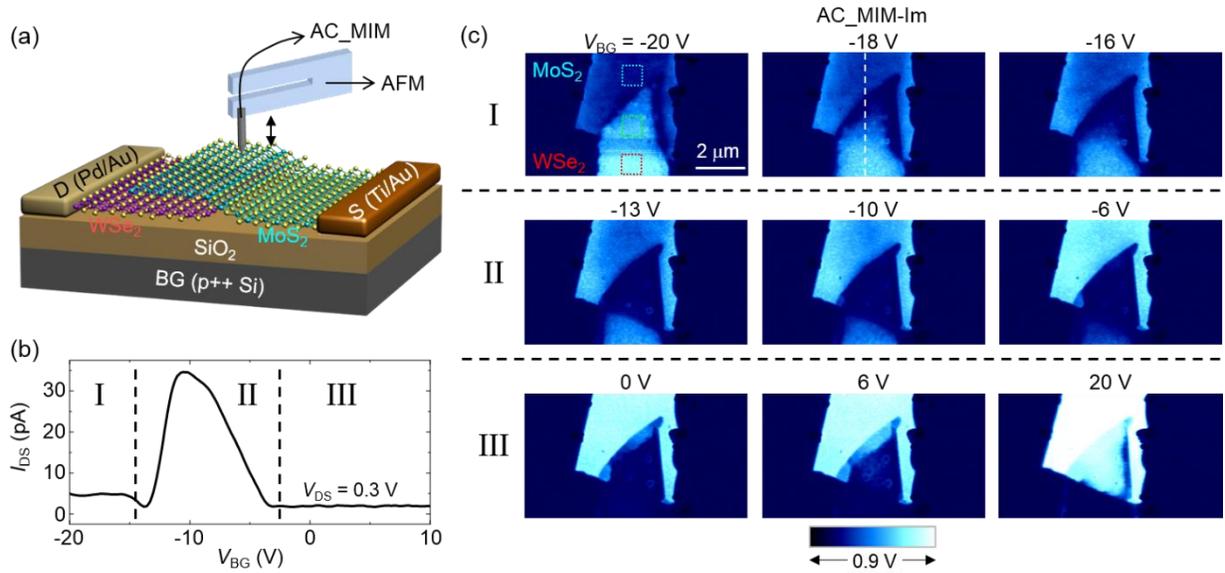

Figure 2. (a) Schematic of the experimental setup, showing the TF-based MIM and the heterostructure transistor. (b) Transfer curve at $V_{DS}$ = 0.3 V. The two dashed lines roughly mark the three regions of the antiambipolar effect. (c) AC_MIM-Im images as a function of $V_{BG}$, separated into three regions according to the transport data. The field of view is the same as that in the AFM image of Fig. 1b.

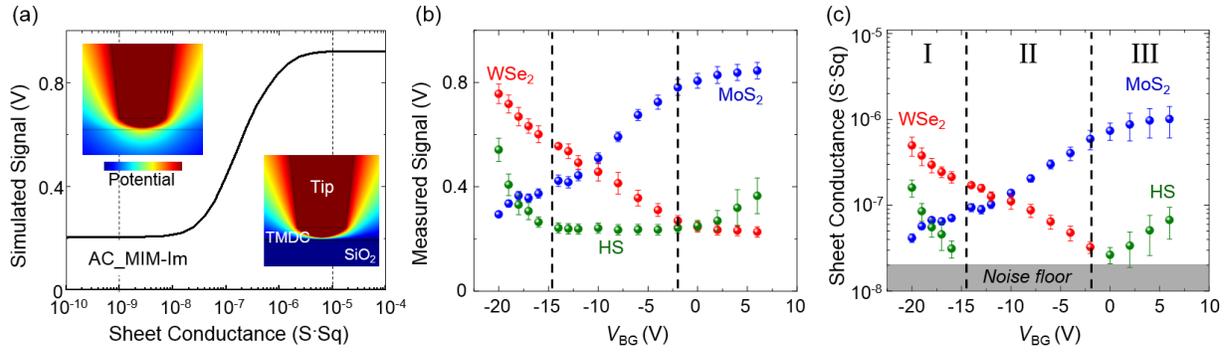

Figure 3. (a) Simulated AC_MIM-Im signal as a function of the 2D sheet conductance $\sigma_{sh}$. The insets show the quasi-static potential distribution at $\sigma_{sh} = 10^{-9}$ and $10^{-5}$ S·sq. (b) Averaged MIM signals inside the dashed squares (blue for $MoS_2$, green for HS, and red for $WSe_2$) in Fig. 2c. (c) Gate dependence of $\sigma_{sh}$ extracted from (a) and (b). The dashed lines again mark the three regions in the transport data.

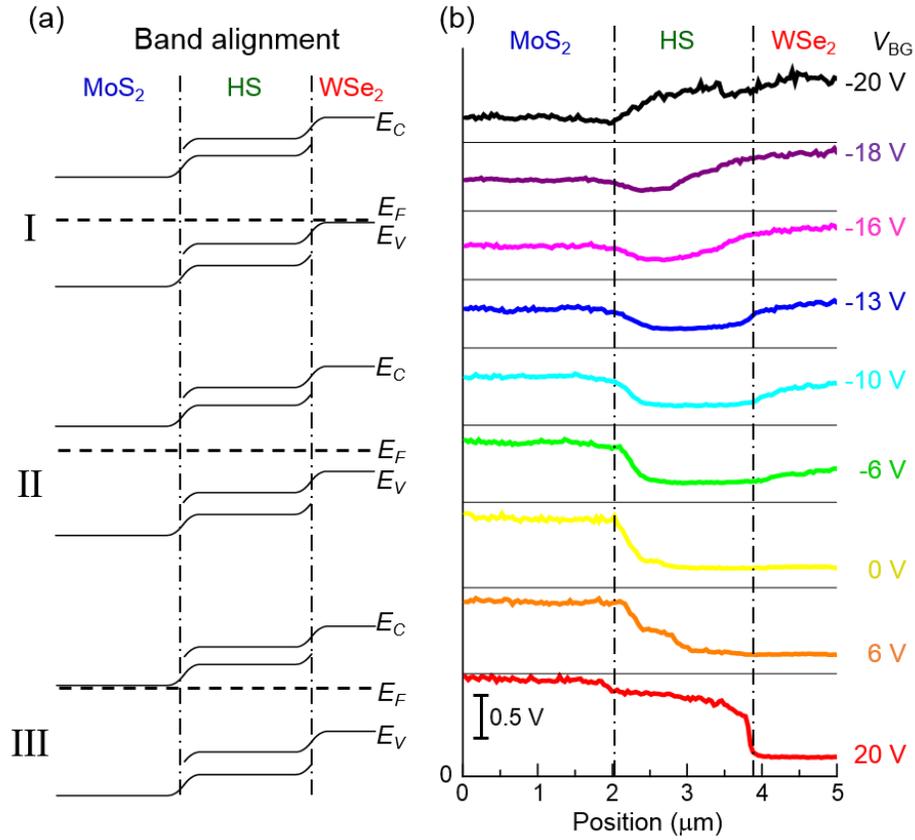

Figure 4. (a) Schematics of the lateral band alignment along the line of $MoS_2$-HS-$WSe_2$ in the three regions at $V_{DS} = 0$ V. $E_C$, $E_F$, and $E_V$ represent the bottom of the conduction band, Fermi level, and the top of the valence band, respectively. (b) Selected MIM line profiles ($V_{DS} = 0$ V) along the white dashed line shown in the $V_{BG} = -18$ V image in Fig. 2c. The zero point baseline is shown at the bottom of each line profile. The gradual change of MIM signals in the length scale of 0.5 μm across the lateral junctions provides a measure of the depletion region width.